\documentclass{PoS}

\title{The Masterclass of particle physics and scientific careers from the point of view of male and female students}

\ShortTitle{Masterclass and scientific careers}

\author{\speaker{Sandra Leone}\\
        INFN Sezione di Pisa\\
        E-mail: \email{sandra.leone@pi.infn.it}}

\abstract{The Masterclass of particle physics is an international outreach activity which provides an opportunity for high-school students
  to discover particle physics. The National Institute of Nuclear Physics (INFN) in Pisa has taken part in this effort since its first year,
  in 2005. The Masterclass has become a point of reference for the high schools of the Tuscan area around Pisa. Each year more than a hundred
  students come to our research center for a day. They listen to lectures, perform measurements on real data and finally they join the participants
  from the other institutes in a video conference, to discuss their results. At the end of the day a questionnaire is given to the students to
  assess if the Masterclass met a positive response. Together with specific questions about the various activities they took part in during the day,
  we ask them if they would like to become a scientist. They are offered 15 possible motivations for a ``yes'' or a ``no''  to choose from.
  The data collected during the years have been analysed from a gender perspective. Attracting female students to science and technology-related
  careers is a very real issue in the European countries. With this study we tried to investigate if male and female students have a different
  perception of scientific careers. At the end, we would like to be able to provide hints on how to intervene to correct the path that seems
  to naturally bring male students towards STEM disciplines (science, technology, engineering, and mathematics) and reject female students from them. }

\FullConference{38th International Conference on High Energy Physics\\
		3-10 August 2016\\
		Chicago, USA}

\begin{document}

\section{Introduction}
The International Masterclasses for Particle Physics (MC) give students the opportunity to be particle physicists for a day~\cite{MC}.
Each year in spring high school students  and their teachers spend one day in reasearch
institutes and universities around the world. They first attend introductory lectures about particle physics
(on the standard model of elementary particles, accelerators and detectors), then they work as scientists, making measurements on real data
collected at CERN by the LHC experiments. At the end of their research day they experience the international aspect of real collaborations in particle
physics, by presenting their findings in a video linkup with CERN or Fermilab and student groups in other participating countries.

The Pisa unit of the National Institute for Nuclear Physics joined the MC since the first year, in 2005 (World Year of Physics)~\cite{PisaMC}.
Each year more than a hundred students 18-19 years old attending the last year (the fifth one)  of high school come to our
institute. They are selected by their schools, taking into account their expression of interest for the initiative and the previous year grades; in addition,
since a few years we ask teachers to reflect the gender distribution of the school in the list of selected students.

At the end of the videoconference a questionnaire is given to the students to assess if the Masterclass met a positive response. Approximately 80\% of the
students taking part to the Masterclass fill the questionnaire.
Together with specific questions about the various activities they attended during the day, we ask them if they would like to become a scientist.
The data collected since 2010 have been analyzed from a gender perspective. About 500 students filled the questionnaire, 300 male and 200 female
students.

\section{Analysis of the questionnaire: general part}

We ask the students several questions related to the various aspects of the Masterclass: {\it were the lectures understandable? was your physics background
adequate? was the measurement fun? was the videoconference easy to follow?}
Then we ask them more general questions: {\it were the Masterclass topics interesting?  was the Masterclass  helpful to better understand
  what physics is and for the choise of your future studies?  after taking part to the Masterclass,
  is your interest for physics  less, equal, or
more than before? is it  worth to participate to a particle physics Masterclass?}

Fig. 1 shows an example of the answers to some of the questions, in blue for male students, in red for female students.
One can see that the distribution of answers is very similar, for male and female students. Fig. 2 (left)  shows the only question for which we get a
different distribution of the answers: {\it are you interested in physics outside school?}
A similar pattern was already observed in a very preliminary study performed on a smaller number of questionnaire in 2010~\cite{giorgio}.

\begin{figure*} [ht]
\begin{center}
  \includegraphics[width=0.46\textwidth]{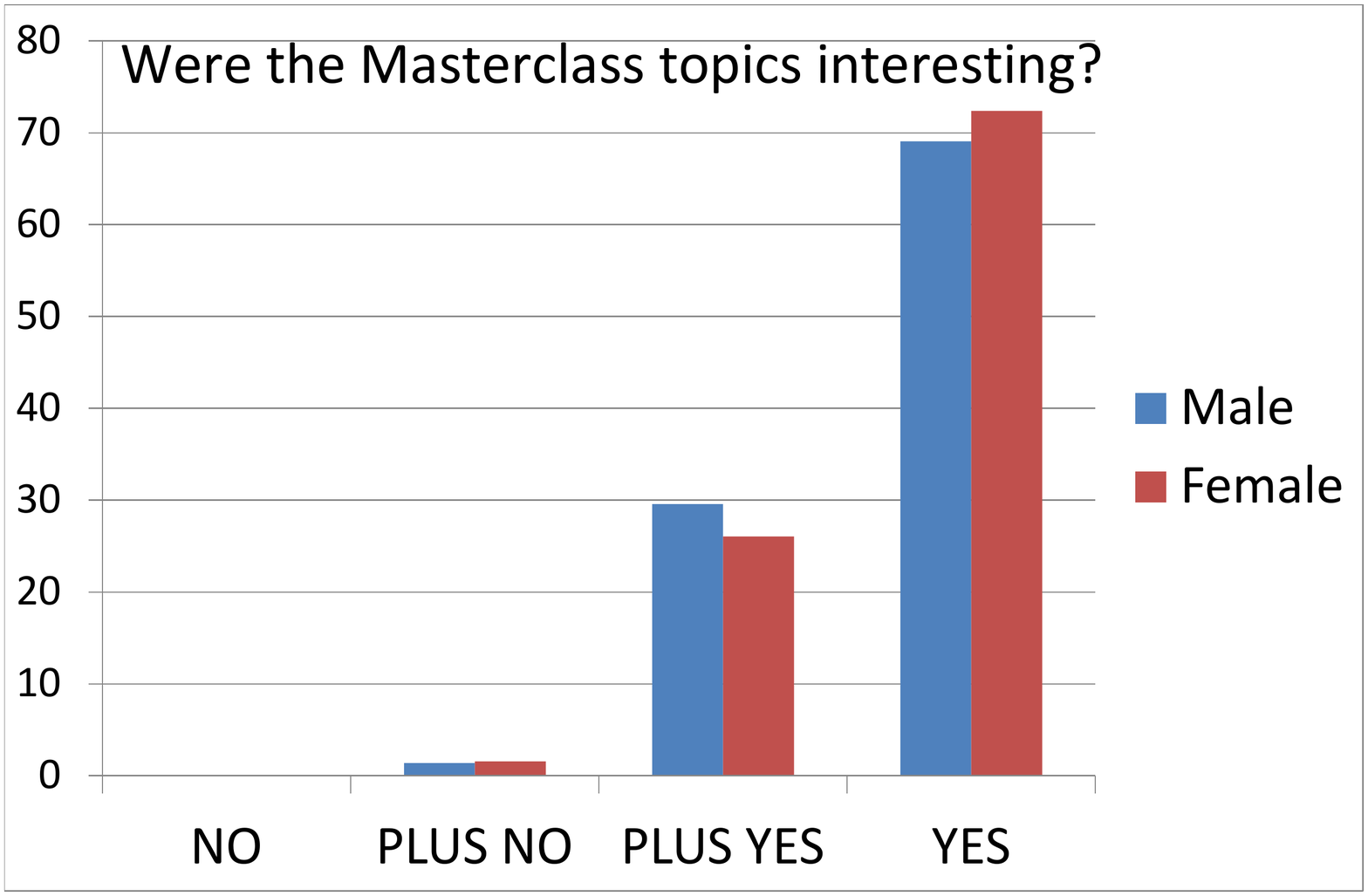}
  \includegraphics[width=0.46\textwidth]{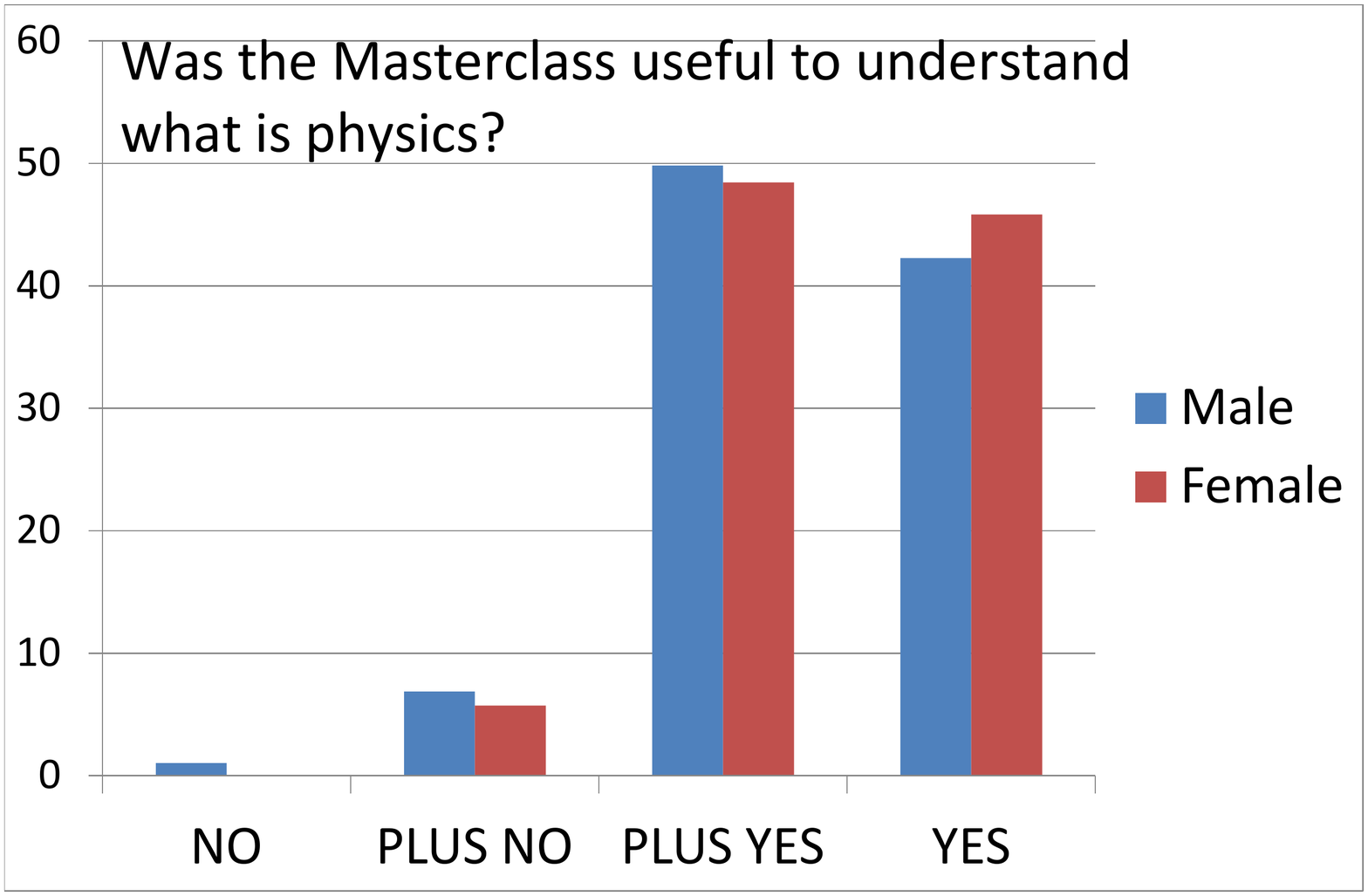}
  \includegraphics[width=0.46\textwidth]{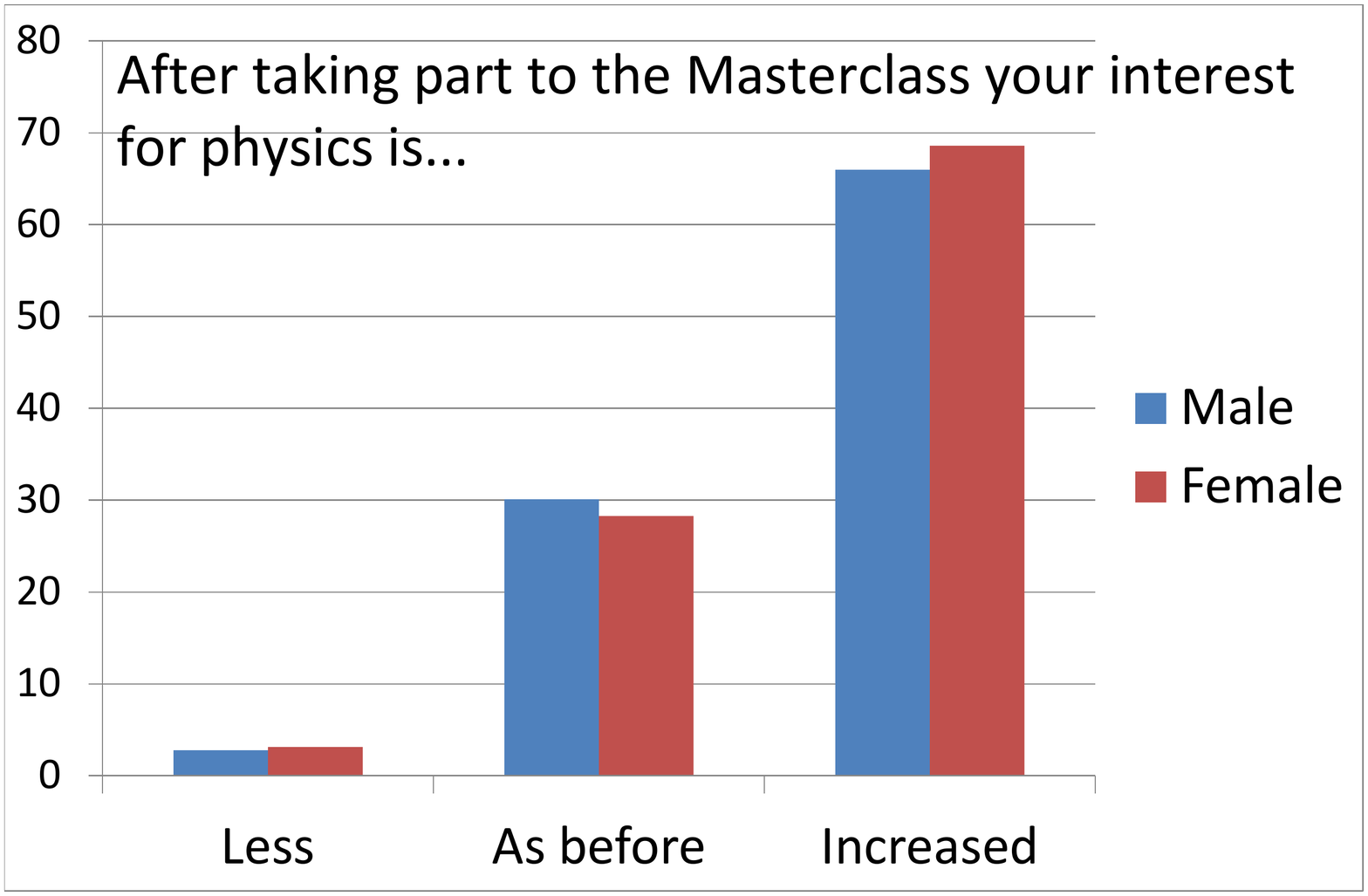}
  \includegraphics[width=0.46\textwidth]{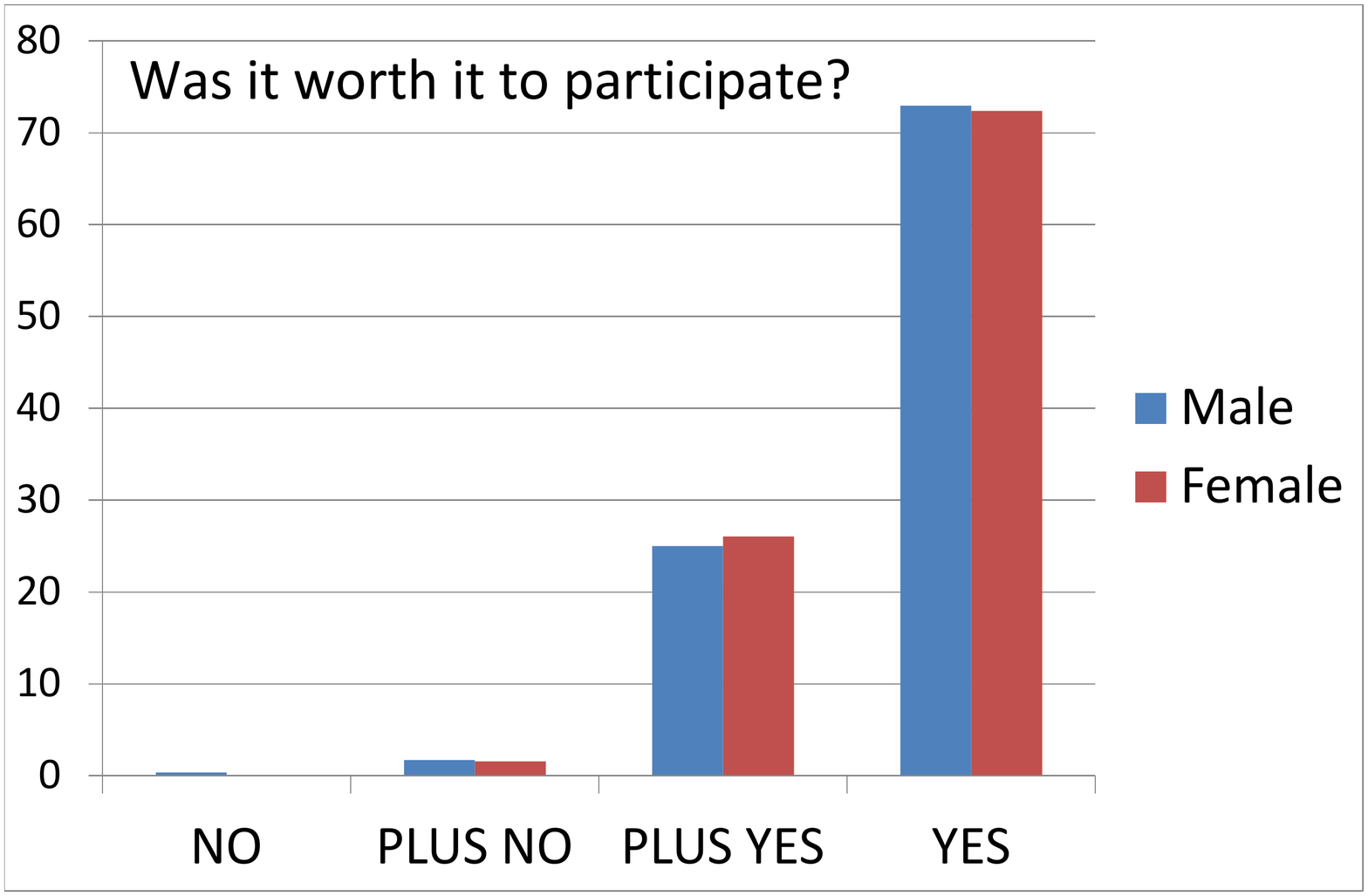}
  \caption{Distribution (in \%) of some of the answers given by male and female students.}
\end{center}
\end{figure*}


\begin{figure*} [!ht]
  \centering
    \includegraphics[width=0.48\textwidth]{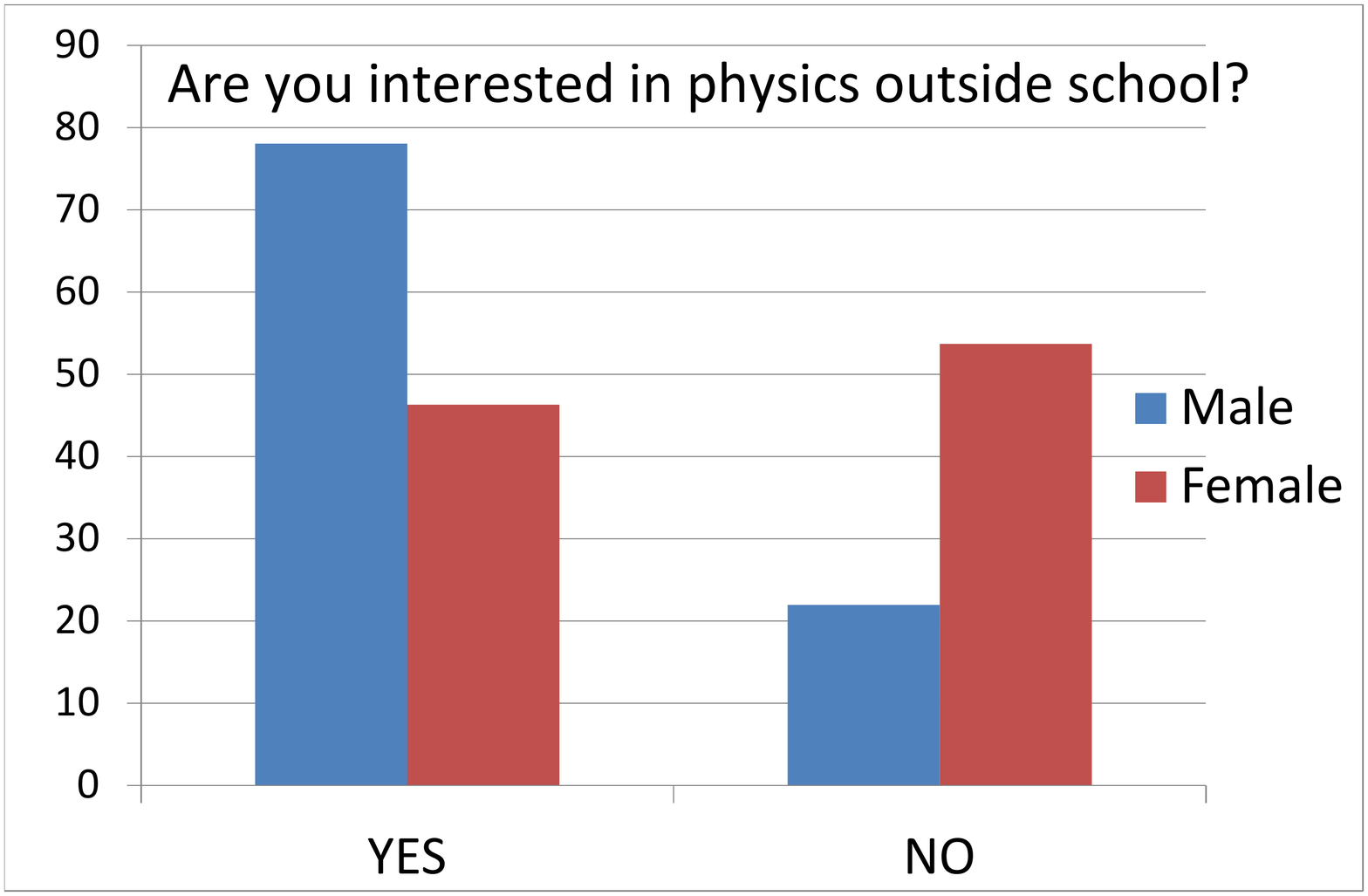}
    \includegraphics[width=0.48\textwidth]{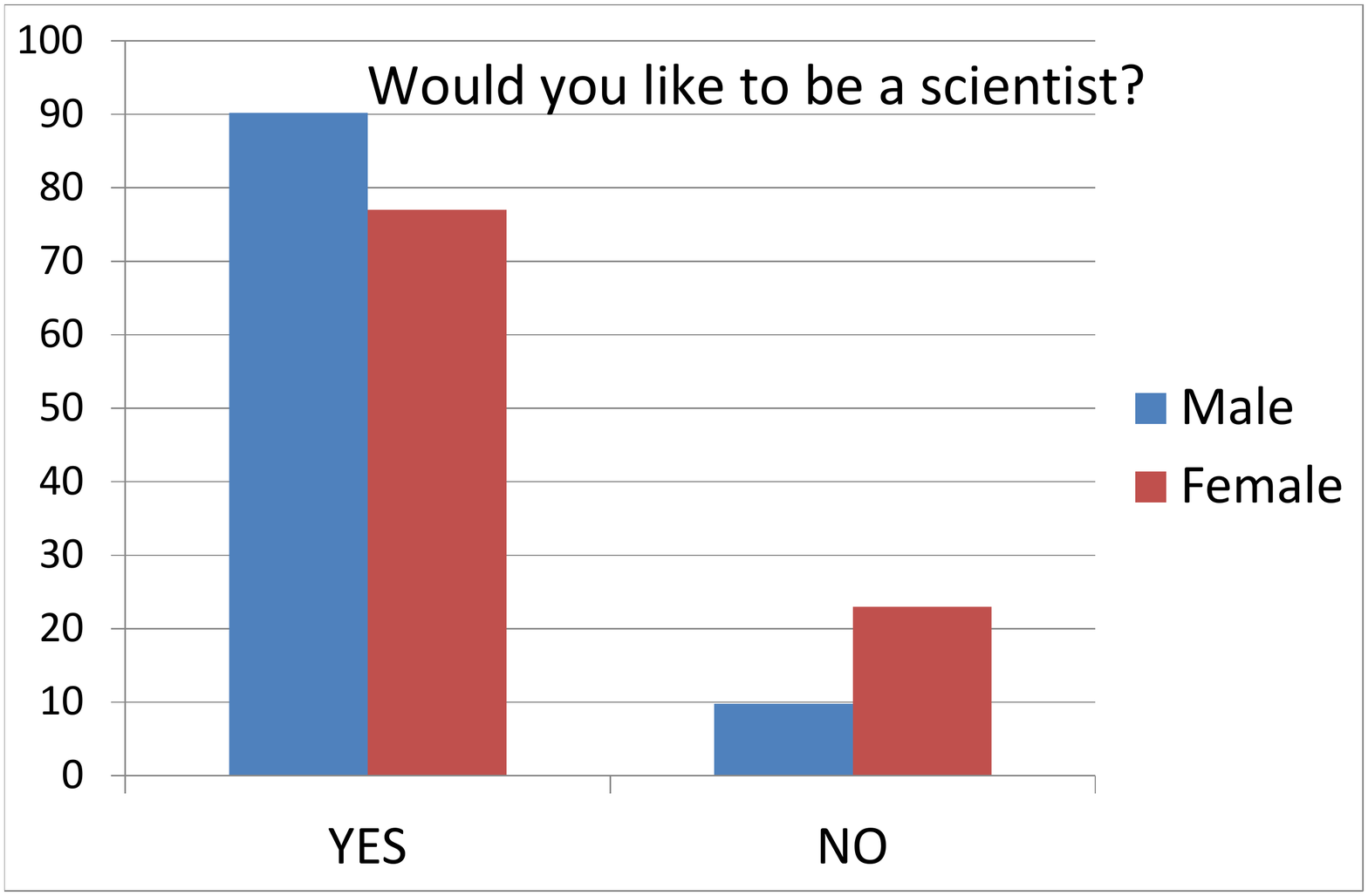}
    \caption{Left: distribution (in \%) of the answer to the question: are you interested in physics outside school? A significant difference between male
      and female students is seen. Right: answer to the question: would you like to be a scientist?}
\end{figure*}

\begin{figure*} [!ht]
  \centering
  \includegraphics[width=0.48\textwidth]{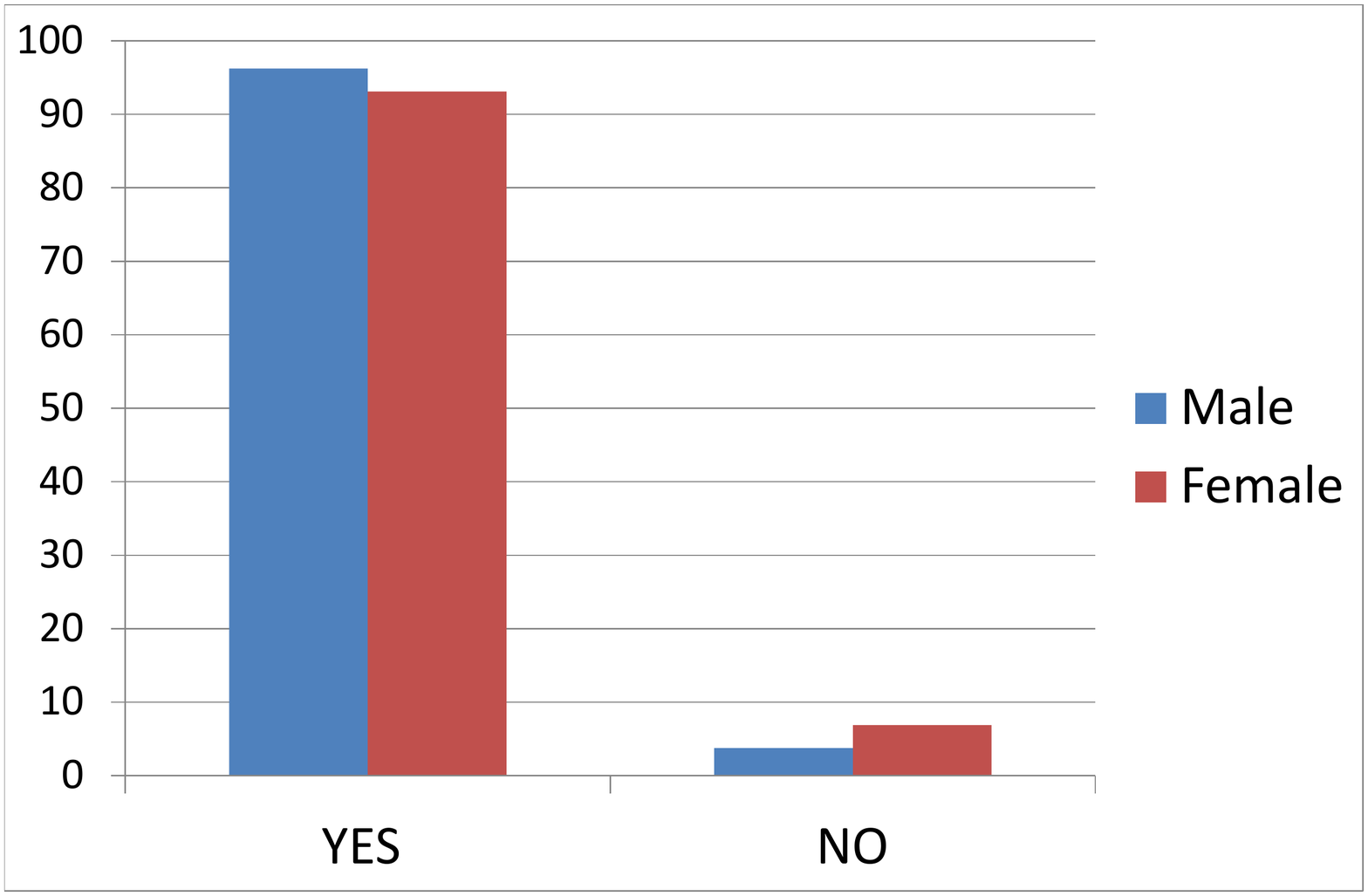}
  \includegraphics[width=0.48\textwidth]{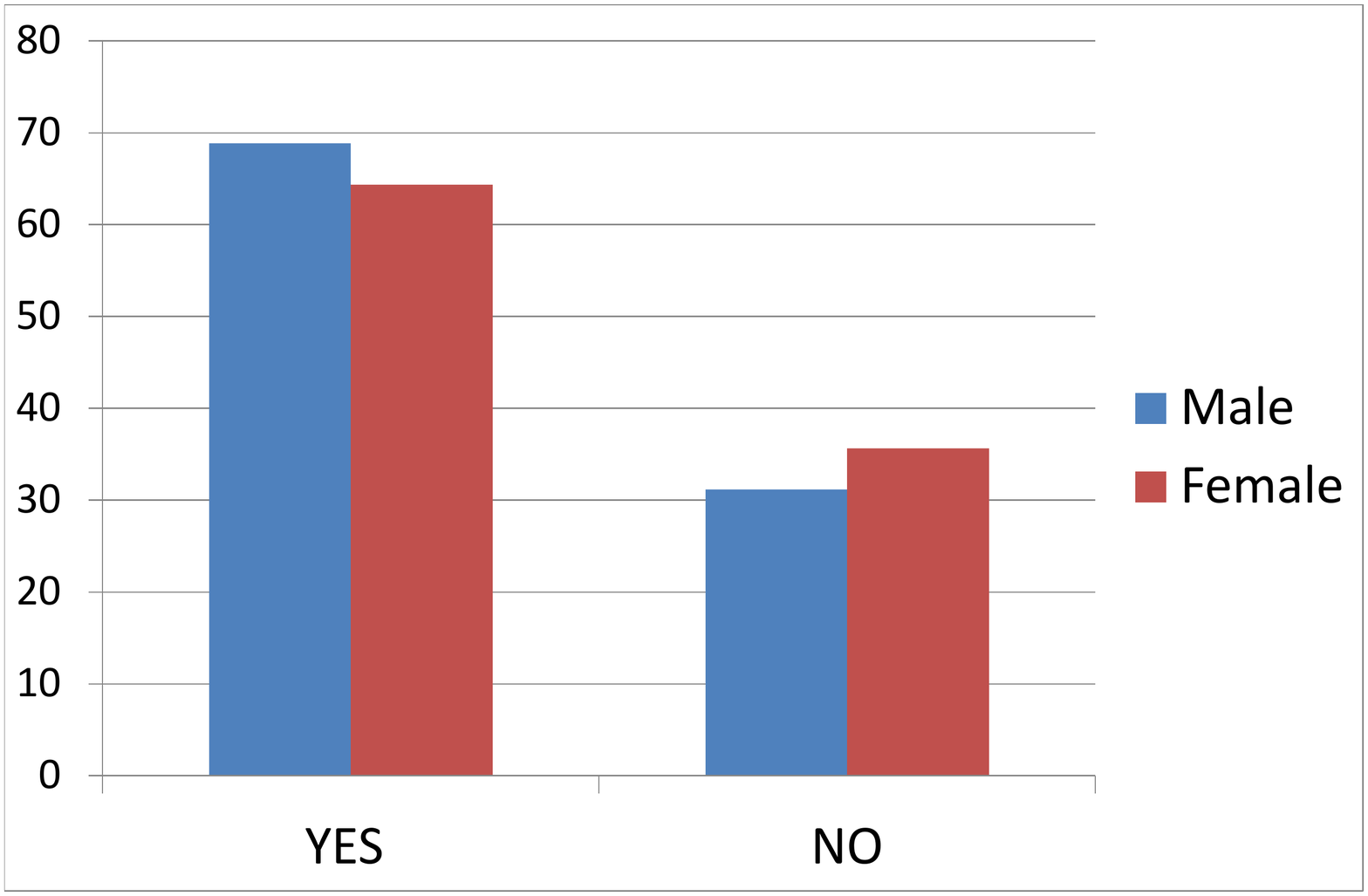}
    \caption{Distribution (in \%) of the answers to the question: would you like to be a scientist? on the left (right) for students interested (not interested)
      in physics outside school.}
\end{figure*}

\section{Analysis of the questionnaire: would you like to be a scientist?}

Finally, we ask the students: {\it would you like to work or do research in a STEM (physics, technology, engeneering, and mathematics)
  discipline?} The distribution of their answers is shown in fig. 2 (right).
A certain difference between male and female answers is seen.

We divided the sample in students who declared to be (not to be) interested in physics outside school, and their answer to the previous question
is shown in fig. 3 left (right). Now the two distributions are very similar, for male and female students.

The students are offered many options to choose from, to motivate their choice, and are asked to select up to a maximum of five reasons
for a ``yes'' or a ``no'' among the ones listed here.

Yes because:
\vspace{-0.3cm}
\begin{itemize}
\setlength\itemsep{-0.4em}
\item  It's s easy to find a job;
  \item  I have a talent for science;
  \item  I see myself as a scientist;
  \item I like science;
  \item I like to do things that are considered difficult;
  \item I like the idea of studying the mysteries of the universe and finding answers to new questions;
  \item I'm not scared by the idea of working in a lab, without regular meals and hours;
  \item One can make a lot of money in science;
  \item It's a field where one can travel a lot;
  \item The choice of career has a high priority in my life;
  \item It would make my life more interesting;
  \item I'm not scared by the prospects of an all-encompassing job;
  \item I deeply admire scientists and consider them a role model;
  \item My teachers are encouraging and are advising me to undertake a scientific career;
    \item My family is encouraging me and would be very happy if I were to choose a scientific career.
\end{itemize}

No, because:
\vspace{-0.3cm}
\begin{itemize}
  \setlength\itemsep{-0.4em}
\item  It's difficult to find a job;
\item I have no talent for science;
\item I cannot see myself as a scientist;
\item I don't like science;
\item Scientific disciplines are too difficult;
\item One has to study too much;
\item I would like to do more useful work;
\item Working in a lab without regular meals and hours is not for me;
\item I put my personal interests first;
\item I don't want to sacrifice my personal life for my career;
\item I aspire to a normal life;
\item I'm scared by the prospects of an all-encompassing job: I want to have time for myself;
\item There aren't scientists who I consider as a model;
\item My teachers are discouraging me;
\item My family is discouraging me.
\end{itemize}

\begin{figure*} [!ht]
  \centering
  \includegraphics[width=1.1\textwidth]{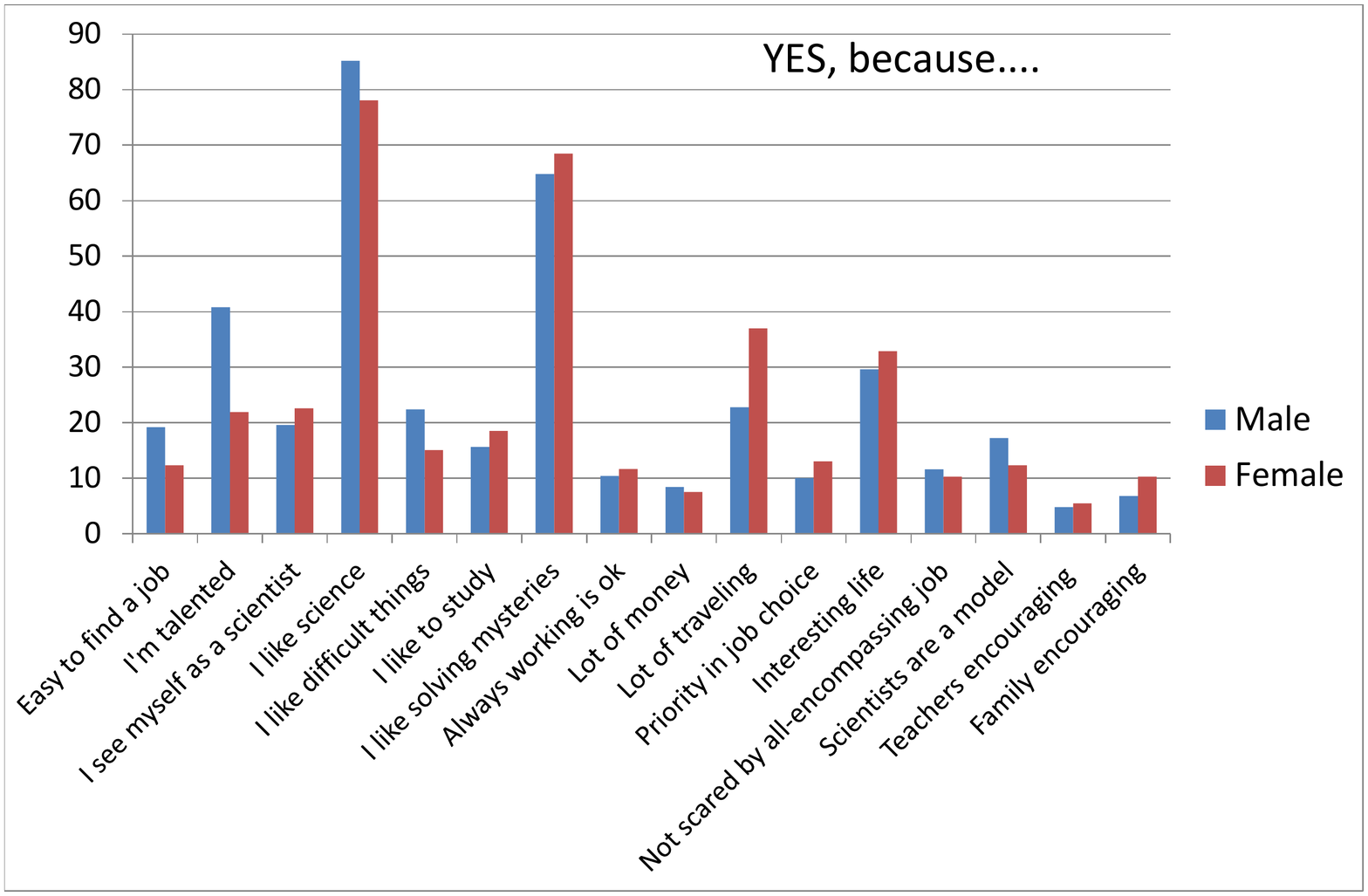}
\vspace{-1.3cm}
  \caption{Distribution (in \%) of the motivations for willing to be a scientist.}
\end{figure*}

From the distribution of the ``yes'' motivations, one can notice that more male (about 40\%) than female (about 20\%) students think that
they have a talent for science. On the other hand,
more female (about 37\%) than male (about 23\%) students are attracted by the idea of traveling.

The interpretation of the ``no'' distribution is affected by large statistical uncertainties, because
only about 70 students answered ``no''. However, it is interesting to notice that, among them,
65\% of female
students feel that they have no talent for science (compared to 40\% of male),
and a few of them are discouraged by family
(while no male student is). In addition, 55\% of male students are afraid that in science they'll not have enough
time for themselves (compared to 7\% of female students).

\begin{figure*} [!ht]
  \centering
  \includegraphics[width=1.1\textwidth]{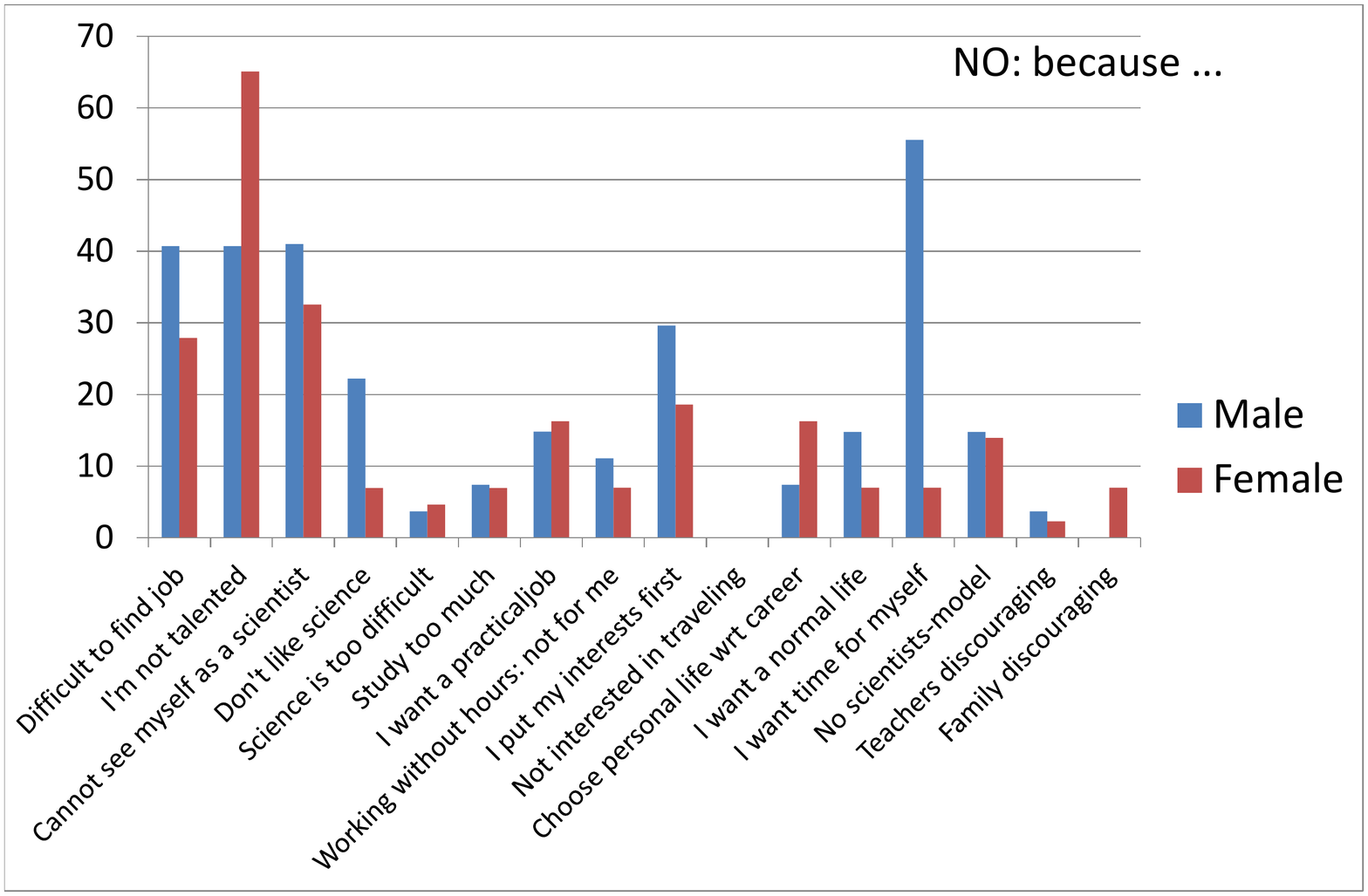}
\vspace{-1.3cm}
  \caption{Distribution (in \%) of the motivation for not willing to be a scientist.}
\end{figure*}

\section{Conclusion}

We present a preliminary analysis of the answers to about 500 questionnaires filled by students attending the Masterclass
of particle physics in Pisa from 2010
to 2016. Looking for differences in answers from male and female students, we notice that almost 80\% of male students declare to be interested
in physics outside school, compared to 46\% of female students. About 90\% of male students say that they would like to work in a STEM
discipline, compared to about 77\% of female students.

We plan to continue to distribute this questionnaire to students attending the Masterclass of particle
physics in Pisa and collect more data.  In addition, we asked the physics teachers to propose the general section of the
questionnaire concerning scientific careers also to students who will not attend the Masterclass. 
This will provide a control sample including students not as good as the ones coming to the Masterclass
and not necessarily interested
in science as a career. We aim to better understand in which respect male students are more interested in physics outside school
than female students.
At the end, we would like to obtain hints on how to intervene to correct the path that seems to naturally bring male students towards STEM disciplines
and reject female students from them.

\end{document}